# Discovery of a two-dimensional topological insulator in SiTe

Yandong Ma[†,⊥]*, Liangzhi Kou[‡], Ying Dai[§], and Thomas Heine[†,⊥]*

[†] Wilhelm-Ostwald-Institut für Physikalische und Theoretische Chemie, Universität Leipzig, Linnéstr. 2, 04103 Leipzig, Germany

[⊥] Department of Physics and Earth Sciences, Jacobs University Bremen, Campus Ring 1, 28759 Bremen, Germany

[‡] School of Chemistry, Physics and Mechanical Engineering Faculty, Queensland University of Technology, Garden Point Campus, QLD 4001, Brisbane, Australia

[§] School of Physics, Shandong University, Shandanan Str. 27, 250100 Jinan, People's Republic of China

*Corresponding author: myd1987@gmail.com (Y.M.); thomas.heine@uni-leipzig.de (T.H.)

Two-dimensional (2D) topological insulators (TIs), a new state of quantum matter, are promising for achieving the low-power-consuming electronic devices owning to the remarkable robustness of their conducting edge states against backscattering. Currently, the major challenge to further studies and possible applications is the lack of suitable materials, which should be with high feasibility of fabrication and sizeable nontrivial gaps. Here, we demonstrate through first-principles calculations that SiTe 2D crystal is a promising 2D TI with a sizeable nontrivial gap of 0.220 eV. This material is dynamically and thermally stable. Most importantly, it could be easily exfoliated from its three-dimensional superlattice due to the weakly bonded layered structure. Moreover, strain engineering can effectively control its nontrivial gap and even induce a topological phase transition. Our results provide a realistic candidate for experimental explorations and potential applications of 2D TIs.

TOC Figure

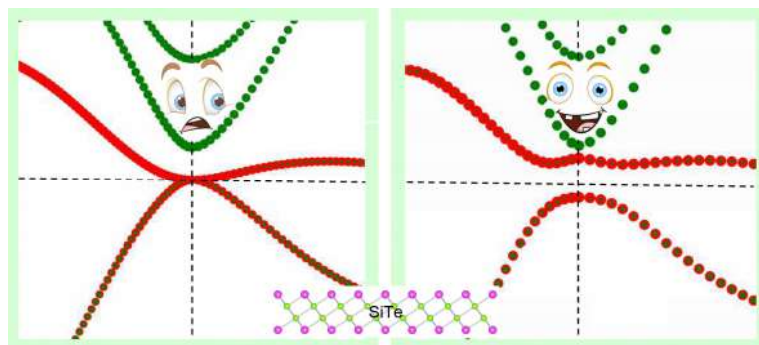

# I. Introduction

Two-dimensional (2D) topological insulators (TIs), also known as quantum spin Hall (QSH) insulators, are new states of quantum matter characterized by the conducting edge states inside the insulting bulk gap as protected by time-reversal symmetry [1,2]. Charge carriers in such edge states are helical Dirac Fermions, which are immune to nonmagnetic scattering and exhibit dissipationless transport, and hence are expected to lead to fascinating applications in energy-efficient electronic devices and spintronics [3]. So far, the experimental verifications of 2D TIs in real systems are limited to the HgTe/CdTe [4] and InAs/GaSb [5] quantum-wells. Moreover, extreme conditions, e.g., precisely controlled molecular-beam epitaxy (MBE) and ultralow temperature (due to their small bulk gap of the order of meV), are required, thus greatly obstructing the exploitation and application of QSH effect. Therefore, there is a great interest in searching for suitable 2D TI materials with sizeable band gaps and characteristics of easy fabrication. Recently, a variety of 2D materials have been predicted to harbor the 2D TI phase, such as oxygen functionalized MXene [6], arsenene [7], $Ni_3C_{12}S_{12}$ monolayer [8], 2D transition-metal halide [9], 2D transition metal dichalcogenides [10-12], III-Bi monolayers [13], $TiS_{2-x}Te_x$ monolayers [14], Bi bilayer [15], BiF thin film [16], functionalized Bi/Sb monolayers [17-19], $Bi_4Br_4$ 2D crystals [20], $ZrTe_5$/$HfTe_5$ 2D crystals [21], functionalized Ge/Sn thin films [22-24], and modified phosphorene [25]. Unfortunately, most of these theoretically proposed structures do not naturally exist and remain to be made in experiments where the requirement of precisely controlled techniques would be a challenge.

Among these systems, the $Bi_4Br_4$ and $ZrTe_5$/$HfTe_5$ 2D crystals seem to be very promising because they can be cleaved from the experimentally existed three-dimensional (3D) layered materials and their nontrivial gaps are sizeable, though their atomic structures might be too complicated. Regarding the IV-VI 2D materials, they are also known to hold great potential for the investigation of nontrivial topological properties and there are several theoretical proposals available in literature, e.g., 2D topological crystalline insulators (TCIs) in rocksalt 2D crystals [26-28]. However, these hypothetical rocksalt IV-VI 2D structures also suffer from structural instability and high surface chemical activity and have yet to be synthesized in experiments. Recently, the hexagonal IV-VI (SiTe, GeTe) 2D building blocks with a simple atomic structure were experimentally confirmed in the IV-VI/$Sb_2Te_3$ superlattices in which the VI-IV-IV-VI and Te-Sb-Te-Sb-Te layers are connected by van der Waals (vdW) interactions [29-31]. Such weak inter-slab interaction allows that the hexagonal SiTe and GeTe 2D crystals can be easily obtained by mechanical exfoliation from these layered superlattices akin to fabricating graphene from graphite.

Here, on the basis of density functional theory (DFT), we demonstrate that the SiTe 2D crystal with a hexagonal structure is a 2D TI with a sizeable bulk gap of 0.220 eV, without any additional turning. And the calculated $Z_2$ invariant and edge states provide direct evidences for its nontrivial topological properties. This sizeable-gap 2D TI is an ideal candidate for further experimental studies on QSH effect because it has simple stoichiometric ratio and crystal structure, and most importantly, it is can be naturally obtained by cleaving from its experimentally existed superlattice. The phonon spectrum calculations and *ab initio* molecular dynamic (MD) simulations further confirm that this 2D crystal can be dynamically and thermally stable. Moreover, a precise control of its topological state can be achieved by applying strain. Our works thus represent a major step forward in realistic applications of QSH effect.

## II. Computational Methods

First-principles calculations based on DFT are carried out by using projector augmented wave (PAW) method as implemented in the plane-wave code VASP [32-34]. For structural relaxations and band structure calculations, generalized gradient approximation with Perdew-Burke-Ernzerhof (PBE) parametrization [35] is employed to simulate the electronic exchange-correction effect. To overcome the problem of band-gap underestimation in semi-local exchange-correlation functionals, Heyd-Scuseria-Ernzerhof (HSE06) hybrid functional [36] is also used for calculating the band structures. If potential is unspecified, the results are based on PBE. A cutoff energy of 500 eV is used throughout all calculations. Structures are fully relaxed until residual forces on each atom less than 0.01 eV/Å. Brillouin zone (BZ) integration is performed with the k-point meshes [37] of 15×15×1. The vacuum space is 20 Å to ensure that interactions between layers are negligible. Spin-orbit coupling (SOC) is included at the second variational step using scalar-relativistic eigenfunctions as a basis. The phonon calculations are carried out using Phonopy program implementing density functional perturbation theory (DFPT) method [38,39]. The *ab initio* MD simulations are performed by adopting the canonical ensemble with a Nosé thermostat.

## III. Results and Discussion

The optimized atomic structure of SiTe 2D crystal is shown in **Figure 1(a)**. This structure possesses a hexagonal Bravais lattice and a *P-3m1* symmetry with an inversion center located at the center of two nonequivalent Si atoms. Within each unit cell [marked by the dashed line in **Figure 1(a)**], it contains two 6-coordinated Si atoms and two 3-coordinated Te atoms. The surfaces of SiTe 2D crystal are basically terminated by the 3-coordinated Te atoms, just like the case of $Bi_2Te_3$ family, and thus they are chemically inert. In fact, by replacing the internal Bi-Te-Bi layer with the Si-Si layer in $Bi_2Te_3$ sheet, the geometry of SiTe 2D crystal can be realized. For more details about the structure, please see supporting information. Both the stoichiometric ratio and crystal structure of SiTe 2D crystal are relatively simple, which would be beneficial to its mechanical exfoliation from the layered superlattice and further applications.

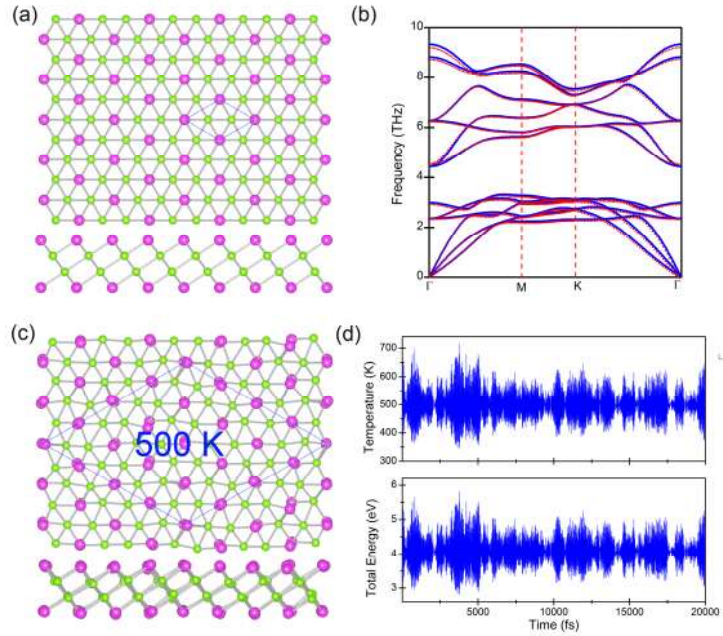

**Figure 1**. (a) Top and side views of the atomic structure of SiTe 2D crystal, with dashed line marking the unit cell. (b) Phonon spectra of SiTe 2D crystal calculated with (red) and without (blue) SOC. (c) Top and side views of the snapshot of SiTe 2D crystal at the end of MD simulation at 500 K; dashed line indicates the supercell used in the simulation. (d) Changes of temperature and energy with time obtained from MD simulation of SiTe 2D crystal at 500 K. The green and pink balls represent Si and Te atoms, respectively.

The structurally stability of SiTe 2D crystal can first be understood though analyzing its phonon spectrum. The phonon band dispersion curves are shown in **Figure 1(b)**. It can be clearly seen that the obtained phonon dispersions without (blue lines) and with (red lines) considering SOC agree well with each other: all the branches have positive frequencies without any imaginary phonon modes in the entire BZ, thus confirming the dynamic stability of this 2D crystal. The thermal stability of SiTe 2D crystal is further investigated by performing *ab initio* MD simulations. To approach the true stability, we adopt a relatively large supercell consisting of 4×4 repeated unit cells [marked by dashed line in **Figure 1(c)**] to simulate the 2D lattice. After heating at 300 K and 500 K for 20 ps with a time step of 1 fs, we find no structure reconstruction in both of the cases, suggesting that SiTe 2D crystal is thermally stable at least up to 500 K. The snapshots of atomic configurations of SiTe 2D crystal at the end of MD simulations at 500 K and 300 K are presented in **Figure 1(c)** and **Figure S1(a)**, respectively, and the corresponding fluctuations of energy and temperature with time during the simulations are shown in **Figure 1(d)** and **Figure S1(b)**. The stability analysis discussed above makes us more confident to believe that such a SiTe 2D building block can form a stable 2D crystal.

The PBE band structures of SiTe 2D crystal without and with SOC are shown in **Figure 2(a)**. In the vicinity of Fermi level, the valence and conduction bands are mainly contributed by Si/Te-$p_{x,y}$ orbitals. Without including SOC, the valence and conduction bands of SiTe 2D crystal exhibit a parabolic character and the valence band maximum (VBM) and conduction band minimum (CBM) degenerate at Γ point with Fermi level locating exactly at the degenerate point. The 3D plot of band dispersions of SiTe 2D crystal around Fermi level is shown in **Figure 2(c)**. Therefore, in the absence of SOC, SiTe 2D crystal is a gapless semiconductor. After turning on SOC, the degenerate point is lifted and Fermi level separates the valence and conduction bands, leading to a sizeable global band gap of 129 meV, as shown in **Figure 2(a)**; and this band gap is related to the strong SOC strength within $p_{x,y}$ orbitals, which is similar to cases of functionalized Ge/Sn thin films [22-24]. SiTe 2D crystal thus becomes an insulator under SOC. Such a SOC-induced transition from a gapless-semiconducor to an insulator strongly suggests that SiTe 2D crystal is a 2D TI [9,16-19,24]. Additionally, the nearby W-shape conduction band near Γ point is another clear sigh of the existence of nontrivial topological state in SiTe 2D crystal. And more proofs will be given below.

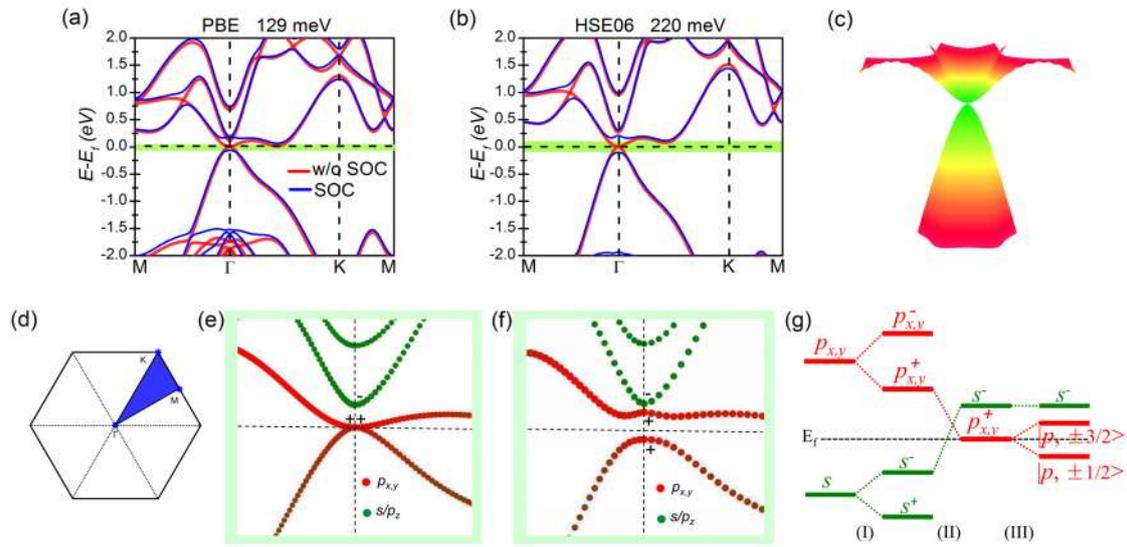

**Figure 2**. Electronic structures of SiTe 2D crystal without and with SOC calculated by using (a) PBE and (b) HSE06 potentials. (c) Band dispersions of SiTe 2D crystal around Fermi level in 2D $k$ space with energy as the third dimension. (d) 2D Brillouin zone of SiTe 2D crystal. Orbital-resolved band dispersions near Fermi level for SiTe 2D crystal (e) without and (f) with SOC calculated by using HSE06 potential; parities of bands at Γ point are labeled by "+" and "-". (g) Schematic diagram of the evolution of energy levels at Γ point for SiTe 2D crystal under (I) chemical bonding, (II) crystal field effect and (III) SOC in sequence.

Since PBE usually underestimates the band gap, we adopt hybrid functional HSE06 to get a more precisely description of the band structures of SiTe 2D crystal, and the corresponding results are plotted in **Figure 2(b)**. The SOC-induced band gap calculated by using HSE06 is found to be 220 meV, significantly larger than the PBE result (129 meV) by 91 meV. Whereas the SOC strength, defined as SOC-induced splitting of the degenerate bands at Γ point, calculated by using PBE (271 meV) and HSE06 (310 meV) are comparable, with a relatively small energy difference of 39 meV. Concerning the band structures of SiTe 2D crystal, the situation is a little special: when turning on SOC, the CBM locates almost at the center of Γ and K points instead of the vicinity of Γ point. From **Figure 2(a)** to **2(b)**, it can been clearly seen that VBM calculated by using HSE06 shifts a little far away from Fermi level as compared with that based on PBE. Therefore, though the difference between SOC strength based on PBE and HSE06 is relatively small, the calculated global band gap based on HSE06 is significantly larger than that of PBE result. Despite the difference in band gap, the main characters of PBE and HSE06 band structures are roughly similar to each other regardless of SOC.

To verify the nontrivial band topology of SiTe 2D crystal, we investigate the topological $Z_2$ invariant following the parity criterion proposed by Fu and Kane [40]. The parities are calculated for each pair of Kramer's degenerate occupied bands at all time-reversal invariant momenta (TRIM) points in the 2D Brillouin zone [see **Figure 2(d)**], one Γ and three M points. The results indicate that SiTe 2D crystal indeed is a 2D TI with $Z_2$=1 (see **Table S1**). Considering its sizeable nontrivial gap of 220 meV, QSH effect can be readily observed in SiTe 2D crystal at room-temperature. The physical manifestation of nontrivial topological invariant is the existence of conducting edge states that are protected by the time-reversal symmetry. To simulate the topological edge states of SiTe 2D crystal, we construct two types of SiTe nanoribbons by cutting the 2D lattices along different directions. The width of SiTe nanoribbons with armchair and linear (from the top view) edges are selected to be 89.7 and 89.3 Å, respectively, to avoid the interactions between two deges. **Figure 3** shows their crystal and band structures. Because each nanoribbon has two symmetric edges (highlighted in green and pink), the edge states from two sides are degenerate in energy. For the nanoribbon with armchair edges, as shown in **Figure 3(a)**, the conducting edge states appear in the band gap and cross linearly with each other at $\bar{\Gamma}$ point. For the nanoribbon with linear edges, although the detailed band dispersions of edge states are different from that with armchair edges, there are still gapless edge states crossing linearly in the band gap at $\bar{\Gamma}$ point; see **Figure 3(c)**. Therefore, if the nontrivial band topology is protected, the conducting edge states can always exist but the details may depend on the shape of edge. These results also confirm the 2D TI state in SiTe 2D crystal.

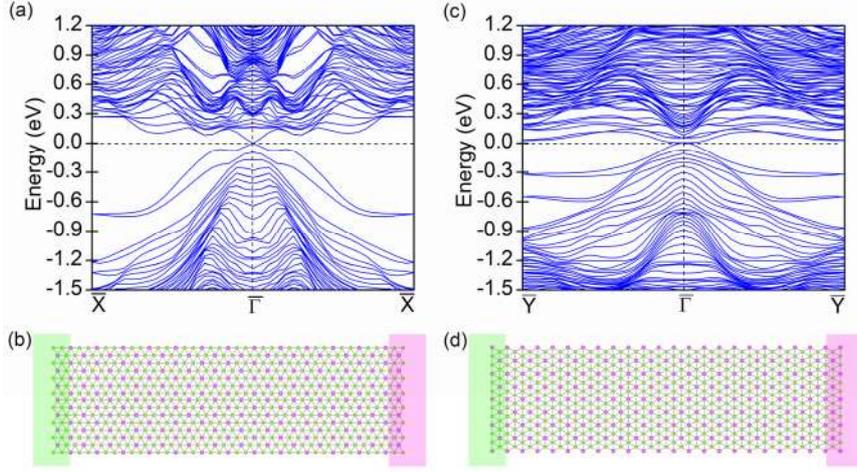

**Figure 3**. Electronic band structures and corresponding crystal structures for SiTe nanoribbon with (a,b) zigzag and (c,d) linear edges. The left and right edges of each nanoribbon are highlighted in green and pink, respectively.

**Figure 2(e)** and **2(f)** show the orbital-resolved band dispersions of SiTe 2D crystal near Fermi level without and with SOC, respectively. Obviously, no band inversion can be observed and thus the nontrivial band order in SiTe 2D crystal is not due to SOC. To uncover the underlying mechanism of the band inversion and topology nature of SiTe 2D crystal, we analyze the orbital evolution at Γ point in detail and the result is shown in **Figure 2(g)**. Since the states around Fermi level are dominated by $s$ and $p_{x,y}$ orbitals, here we focus on these orbitals and neglect other ones. We start from the isolated atomic orbitals and consider the effect of (I) chemical bonding, (II) crystal field and (III) SOC in sequence. At the stage (I), atomic orbitals hybridize to form bonding and antibonding states, which are labeled as $|s^{\pm}\rangle$ and $|p_{x,y}^{\pm}\rangle$ (superscripts ± represents parities of states). Among them, as shown in **Figure 2(g)**, $|s^{-}\rangle$ and $|p_{x,y}^{+}\rangle$ are close to Fermi level, with $|s^{-}\rangle$ locating below $|p_{x,y}^{+}\rangle$. Here the model under chemical bonding effect is estimated by stretching the lattice constant of 150%. When crystal field effect is taken into account in stage (II), namely compressing the structure to equivalent position, band inversion occurs between $|s^{-}\rangle$ and $|p_{x,y}^{+}\rangle$, leading to a nontrivial band order in SiTe 2D crystal. In the inverted band structure, $|s^{-}\rangle$ is unoccupied and degenerate $|p_{x,y}^{+}\rangle$ is half occupied, thereby the systems becomes a gapless semiconductor. After further introducing SOC in stage (III), the degeneracy of the half-occupied $|p_{x,y}^{+}\rangle$ is lifted and an energy gap is created. During this stage, no parity exchange between occupied and unoccupied states is observed, that is, the inverted band order remains unchanged. Consequently, the band inversion in SiTe 2D crystal stems from crystal field effect instead of SOC, and the effect of SOC is to create an energy gap. It is interesting to note that the $s$-$p$ band inversion in SiTe 2D crystal is different from the conventional $s$-$p$ band inversion where $|s^{-}\rangle$ is typically occupied in the inverted band order [22-24].

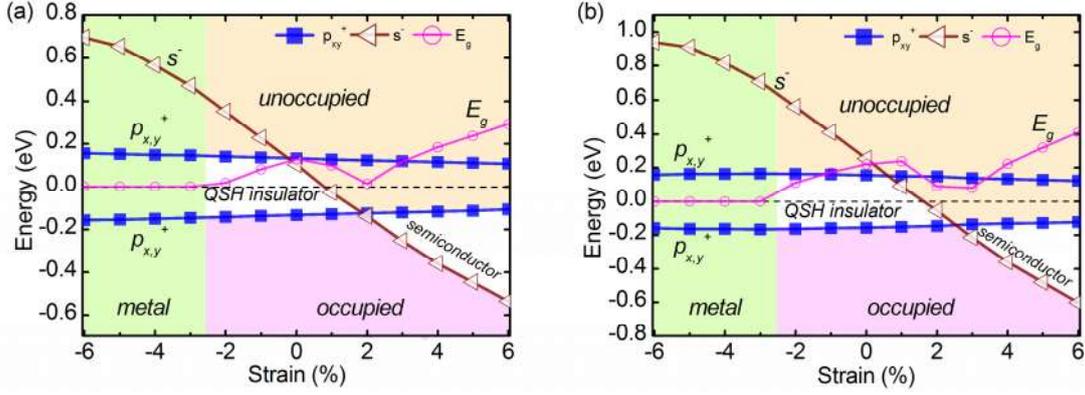

**Figure 4**. The relative positions of $|s^->$ and $|p_{x,y}^+>$ states at Γ point and the global band gap ($E_g$) of SiTe 2D crystal as a function of strain calculated by using (a) PBE and (b) HSE06 potentials.

In what follows, we investigate the effect of external strain on the topological properties of SiTe 2D crystal. The strain range considered here is from -6% to 6%. It should be noted that, for situations under strain in the range from -6% to -3%, they are more about academic interest rather than for practical applications, because currently applying compressive strain is still a challenge in experiment. **Figure 4(b)** shows the relative positions of $|s^->$ and $|p_{x,y}^+>$ states at Γ point and the global band gap ($E_g$) of SiTe 2D crystal as a function of strain calculated by using HSE06, and **Figure S2** shows the corresponding band structures under various strains. Within the strain range from -2% to 2%, the band topology of SiTe 2D crystal remains nontrivial, which is identified by topological $Z_2$ invariant. And the corresponding nontrivial band gap increases from 110 meV to 239 meV and then decreases to 90 meV with increasing strain from -2% to 2%. When decreasing the compressive strain by more than -3%, though $|s^->$ still lies above the upper $|p_{x,y}^+>$, the absence of a global band gap suggests that SiTe 2D crystal transforms into a metal. On the other hand by increasing the stretch strain to 3%, as shown in **Figure 4(b)**, $|s^->$ shifts below the lower $|p_{x,y}^+>$, yielding to a parity exchange between the unoccupied and occupied bands at Γ point, and in turn, inducing a topological phase transition from nontrivial phase to trivial phase. It will be very interesting to observe such topological phase transition experimentally, simply by stretching the sample. For the result calculated by using PBE, as shown in **Figure 4(a)** and **Figure S3**, it shares similar trend with that of HSE06, but with a slight difference in critical strains and global band gaps.

Furthermore, we extend our study to hexagonal GeTe and SnTe 2D crystals, which are counterparts of SiTe 2D crystal. **Figure S4** plots their PBE and HSE06 band structures without and with SOC. We find that, different from SiTe 2D crystals, both GeTe and SnTe 2D crystals turn out to be trivial semiconductors with an indirect band gap. To see whether it is from the changes of their crystal structures, we calculate the band structures of GeTe and SnTe 2D crystals by employing the crystal structure of SiTe 2D crystal without any further relaxation: under this condition, GeTe 2D crystal is still a trivial semiconductor [see **Figure S5(a)** and **(b)**], while SnTe 2D crystal becomes a 2D TI with a nontrivial topological invariant [see **Figure S5(c)** and **(d)**]. Clearly, the absence of topological phases in GeTe and SnTe 2D crystals stems not only from the changes of structures but also from the difference in the intrinsic properties of atoms (such as electronegativity, radii).

## IV. Conclusion

In summary, using first-principles calculations, we identify that SiTe 2D crystal with simple stoichiometric ratio and crystal structure is an ideal 2D TI, presenting a sizeable nontrivial gap of 0.220 eV. The mechanism for its nontrivial band topology is from the band inversion between $|s^->$ and $|p_{x,y}^+>$ states induced by crystal field effect. This material is dynamically and thermally stable. Remarkably, different from most of the previous proposed 2D TIs where the materials do not naturally exist and remain to be synthesized, it can be naturally obtained by cleaving from its experimentally existed superlattice. Besides, a precise control of its band topology can be achieved by applying strain. Considering these merits, we envision that QSH effect can be observed experimentally in SiTe 2D crystal soon, which will greatly advance the practical applications of 2D TIs.

## Supporting Information

The results of MD simulation at 300 K, parities of the occupied bands, PBE and HSE06 electronic structures of SiTe 2D crystal under various strains, and electronic band structures of GeTe and SnTe 2D crystals. This material is available free of charge via the Internet.

## Acknowledgement

Financial support by the European Research Council (ERC, StG 256962) and the National Science foundation of China under Grant 11174180 are gratefully acknowledged.